\documentstyle[12pt]{article}
\begin{document}

\def\del{\partial}
\def\wf{W_\infty^f}
\def\be{\begin{equation}}
\def\ee{\end{equation}}
\def\M{{\widetilde M}}
\def\eq#1{(\ref{#1})}
\def\mbar{\overline{M}}

%\begin{titlepage}
\begin{flushright}
TIFR-TH-94/09
\end{flushright}
\begin{center}
\vspace{3 ex}
{\Large\bf
String Field Theory of Two Dimensional}\\
\vspace{1 ex}
{\Large\bf QCD: A Realization
of $W_\infty$ Algebra}\\
\vspace{8 ex}
Avinash Dhar,  Gautam Mandal and Spenta R. Wadia\\
Tata Institute of Fundamental Research \\
Homi Bhabha Road, Bombay 400 005, INDIA \\
\vspace{15 ex}
\bf Abstract\\
\end{center}
\vspace{2 ex}
We consider the formulation of two dimensional QCD in terms of gauge
invariant bilocal operators (string field) which satisfy a $W_\infty$
algebra. In analogy with our work on the $c=1$ string field theory
we derive an action and associated constraints for the
bilocal field using the method of coadjoint orbits. The $1/N$
perturbation theory around a classical solution that corresponds to
the filled Dirac sea leads to the 'tHooft equation for meson
fluctuations. It is shown that the spectrum of mesons, which are the
higher string modes, transform as a representation of the wedge
subalgebra $W_{\infty+} \otimes W_{\infty-}$. We briefly discuss the
baryon as a stringy solitonic configuration and its characterization
in terms of $W_\infty$ algebra.
\thispagestyle{empty}
\newpage
\noindent{\large\bf Introduction}

\vspace{5 ex}

Gauge theories and string theory have a long standing symbiotic
relationship.  It was the string model that inspired the $1/N$
expansion of non-abelian gauge theories \cite{THOOFT}.  In this work
'tHooft discovered the connection between Feynman diagrams of matrix
valued field theories and two dimensional Riemann surfaces.
Subsequently stringlike equations were derived for Wilson loops in the
large $N$ limit \cite{MMSRW}.

More recently matrix models have been used to define and exactly solve
low dimensional string theories
\cite{DTRS,DOUBLE}.  In particular there has been much
work in studying the $c=1$ string field theory.  This theory has an
exact representation in terms of non-relativistic fermions
\cite{FERMION} and is
characterized by the infinite dimensional Lie algebra $W_\infty$
\cite{WINF}.  An exact
boson representation of this theory can be constructed in terms of the
bilocal field $\Phi (x, y) = \psi (x) \psi^+(y)$ where $\psi(x)$ is the
non-relativistic fermion field.  This was done by constructing the
co-adjoint orbit of $W_\infty$ (characterized by a quadratic
constraint, and the fermion number) and using Kirillov's method to
write down the action\cite{COADJOINT}.
A rigorous basis for this construction was
given by the method of $W_\infty$ coherent states \cite{PATHINTEGRAL}.

In this paper we apply the above method to two dimensional
QCD \cite{BOOK}.  Using gauge invariance we can write the theory
entirely in terms of the gauge invariant bilocal
operator
$M_{\alpha\beta,ij} (x, y) = (1/N) \sum^N_{a=1}
\psi^a_{i\alpha} (x) \psi^{+a}_{j\beta} (y)$, which satisfies the
infinite dimensional algebra
$\wf = W_\infty \otimes U(n_f) $.
This is the string field of two dimensional QCD (see Sec. 1 for
notation).  Gauge invariance
leads to a quadratic constraint on the string field.  This constraint
and the baryon number characterize the co-adjoint orbit of $\wf$
and one can once again construct the action using Kirillov's method
(Sec. 2).
In this formulation, as expected, $N$ (the number of colours) plays the
role of $1/\hbar$.  Bilocal operators in two-dimensional QCD have
been previously considered in \cite{WITTEN} and \cite{KIKKAWA}. More
recently there have been several papers
\cite{GROSSTAYLOR,DOUGLAS,MINAHAN} on the connection between
two-dimensional Yang-Mills theory and string theories.

In this reformulation of two-dimensional QCD as an open string field
theory we get several new insights into both string theory and gauge
theories. In a sense many promises which were not realised in $c=1$
string field theory are fulfilled here%
\footnote{We should point out that the QCD strings considered
here are {\sl a priori} not the same as continuum strings represented
by an integral over world-sheet metrics on smooth surfaces. We make
some remarks about possible connections in Sec. 4.}. For one thing, we
find that there is an infinite tower of physical, stable mesons (in
the large $N$ limit) which form the excitations of the string field
theory and transform into one another as a representation of the group
$W_{\infty+}
\otimes W_{\infty-} \otimes
U(n_f)$ (Sec. 3).
Recall that in contrast in $c=1$ string field theory, which
also formed a representation of $W_\infty$, there was only one physical
particle, the tachyon, along with discrete moduli. It appears that in
the present model the dynamical gauge field which gives rise to
effective interactions between fermions provides a far richer
structure to the theory.

The second remark concerns solitonic configurations \cite{WITTEN}\ in
this model. In the large $N$ limit the string field $M_{\alpha\beta,
ij}(x,y)$ becomes classical and is related very simply to the
self-consistent Hartee-Fock potential in which the fermions move.
Nontrivial classical solutions of the equation of motion for the
string field correspond to different self-consistent Hartree-Fock
potentials which arise out of populating quasi-particle wavefunctions
above the Dirac sea. These correspond to baryons in the theory
(Sec. 4). In the language of the group $W_\infty^f$ these different
classical solutions correspond to different {\em representations} of
the group.  This brings forth the essential point about a {\em stringy
soliton} which, in contrast with solitons in local field theories
which are characterized by discrete or finite dimensional groups, must
be characterized by infinite dimensional groups generated by loop
operators. We also mention that the amplitude for creation of
a soliton-antisoliton pair in this theory goes as \cite{WITTEN}
$e^{-N} \sim e^{-1/\sqrt{\hbar}}$ which is indicative of stringy
nonperturbative effect \cite{SHENKER}.

\vspace{5 ex}

\noindent
{\large\bf 1. Hamiltonian Formulation of 2-dimensional QCD}

\vspace{5 ex}

In this section we review the Hamiltonian formulation of
two-dimensional QCD (QCD$_2$) in the light-cone gauge to set up our
notation (the discussion here is similar to the one in
\cite{KIKKAWA}).  We consider the gauge group $SU(N)$ and the
corresponding gauge fields $A^{ab}_\mu (\mu = 0, 1)$ which are
traceless hermitian matrices.  The fermions are as usual denoted by
$\psi^a_{i\alpha} (x)$, where $a = 1, \ldots, N$ is the colour index,
$i = 1, \ldots, n_f$ is the flavour index and $\alpha = 1, 2$ is the
Dirac index.

The Lagrangian of QCD$_2$ is
\begin{equation}
{\cal L} = -{1 \over 4}~trF_{\mu\nu} F^{\mu\nu} + \bar \psi i D\!\!\! /
\psi + m\bar \psi \psi
\label{one}
\end{equation}
where $F_{\mu_\nu} = \partial_\mu A_\nu - \partial_\nu A_\mu +
i(g/ \sqrt N) [A_\mu,
A_\nu]$, $D\!\!\! / =  \gamma^\mu \left(\partial_\mu +i(g/ \sqrt
N)~A_\mu \right)$ and $\gamma^\mu$ are the gamma matrices in two
dimensions.
We will find it convenient to use light-cone coordinates%
\footnote{Our notation
for light-cone variables is summarized in the Appendix.} in which
the Lagrangian gets rewritten as:
\be
\begin{array}{l}
{\cal L} = 2{\rm tr}F_{+-}^2 + 2\psi_-^\dagger (i\del_+
- (g/ \sqrt{N}) A_+) \psi_-   \\
 ~~~~~~~~~ + 2\psi_+^\dagger (i\del_- -
(g/\sqrt{N}) A_-)
\psi_+ + m (\psi_+^\dagger \psi_- + \psi_-^\dagger \psi_+)
\big)  \\
F_{+-} =  \del_+ A_- - \del_- A_+ +  i(g/ \sqrt{N}) [A_+,A_-]
\end{array}
\label{onea}
\ee
In the following we shall use the light-cone gauge\footnote{The reason
we prefer to work in this gauge is that it is Lorentz-covariant.}
\be
A_+ = {1\over 2} A^-  \equiv {1\over 2} (A_0 + A_1) = 0
\label{two}
\ee
Note that in this gauge we still have symmetry under
local gauge transformations
which depend only on the variable $x^-$. Let us regard $x^+$ as the
``time'' and $x^-$ as ``space''.
To find the hamiltonian we need to go through the usual procedure of
finding out the canonical momenta and the constraints. The constraint
in the gauge sector (Gauss law, eq. (\ref{five}))
comes in the usual fashion by imposing the
requirement that the hamiltonian evolution preserves the gauge (\ref{two}).
In the fermionic sector the use of the light-cone coordinates leads
to an additional constraint coming from the fact that there are
no $x^+$-derivatives on $\psi_+$ in (\ref{onea}); hence the equation
of motion for $\psi_+$, (\ref{fivea})
turns out actually to be a constraint on the
theory. Summarizing, we have the hamiltonian
\begin{equation}
\begin{array}{l}
H \equiv P_+ = \int dx^- \big[ (1/8){\rm tr}~E^2  -(m/2)
\big(\psi^\dagger_-(x)
\psi_+(x) + \psi^\dagger_+(x) \psi_-(x) \big) \big] \\
\left[A_-(x^-,x^+), E(y^-, x^+) \right] = i \delta (x^- - y^-)
\end{array}
\label{four}
\end{equation}
along with the gauss law constraint
\begin{equation}
G^{ab}_- \equiv \del_- E^{ab}  + i{g\over \sqrt{N}}
\left[A_-, E\right]^{ab} -2 {g
\over \sqrt N} \left(\psi^{\dagger b}_-\psi^a_- - {1 \over N}~\delta^{ab}
\psi^{\dagger c}_-\psi^c_- \right) = 0
\label{five}
\end{equation}
and the fermionic constraints
\be
\begin{array}{l}
2(i\del_- - (g/ \sqrt{N}) A_- ) \psi_+ + m\psi_- = 0 \\
2\psi_+^\dagger
(i \!\! \stackrel{\leftarrow}{\del_-} -
(g/\sqrt{N}) A_-) + m\psi_-^\dagger = 0
\end{array}
\label{fivea}
\ee
The fermionic constraint just means that so far as calculation of
correlation functions involving the $\psi_-$'s is concerned we can
forget about this constraint; in correlation functions involving
the $\psi_+$'s we can use the constraint to express them in terms
of $\psi_-$'s. We shall henceforth assume that this has been done.

In the quantum theory the Hamiltonian and the constraint act \break
on the
Schr\"odinger wave functional.  In particular $G^{ab}_- |\Psi\rangle =
0$ expresses the gauge invariance of $|\Psi \rangle$ under local gauge
transformations which are functions of $x^-$.
Equivalently $|\Psi \rangle$ is only a functional of
the orbit $A_-^{[\Omega]} = \Omega A_- \Omega^+ + i\Omega
\partial_-\Omega^+, \psi^{[\Omega]}_- = \Omega\psi_-$ which
can be parametrized by the choice of the gauge
$A_- = 0$.  In this case we can solve for the electric field $E$
from (\ref{five}) and substitute in (\ref{four}) to get
\begin{equation}
\begin{array}{l}
H  =
 (g^2/4 N) \int dx^-dy^- \bigg( \psi^a_{i-} (x^-)
\psi^{\dagger a}_{j-} (y^-) |x^- - y^-|~ \psi^b_{j-}
(y^-)~\psi^{\dagger b}_{i-} (x^-) \\
 ~~~~~~~~~~~ - {1 \over N}~\psi^a_{i-}(x^-) \psi^{\dagger a}_{i-}
(x^-) |x^- - y^-|~\psi^b_{j-} (y^-) \psi^{\dagger b}_{j-} (y^-) \\
 ~~~~~~~~~~~
 - (im^2/4)\int dx^-dy^- {\rm sgn}(x^- - y^-)\psi^a_{i-}(x^-)
\psi^{\dagger a}_{i-}(y^-)    \bigg)
\end{array}
\label{six}
\end{equation}
where $m$ is a cutoff dependent constant that includes the bare
quark mass.

Let us now introduce the gauge invariant string field
\be
M_{\alpha\beta,ij} (x^-,y^-;x^+) =
{1 \over N} \sum^N_{a=1} \psi^a_{i\alpha} (x^-,x^+)
\psi^{\dagger a}_{j\beta} (y^-, x^+)
\label{seven}
\ee
Here $\alpha,\beta$ are spinor indices taking values $+,-$.
Note that $M_{\alpha\beta,ij} (x^-,y^-;x^+)$
is gauge invariant in
$A_- = 0$ gauge.  Its manifestly gauge invariant form is
\begin{equation}
M_{\alpha\beta,ij} (x^-,y^-;x^+) = \psi^a_{i\alpha
} (x^-,x^+) \left(e^{i \int^{y^-}
_{x_-}
A_-(z^-,x^+) dz^-} \right)_{ab} \psi^{\dagger
b}_{i\alpha} (y^-,x^+)
\label{eight}
\end{equation}
The hamiltonian (\ref{six})
can be expressed entirely in terms of the string field $M
_{\alpha\beta,ij}(x^-,y^-;x^+)$,
\begin{equation}
\begin{array}{l}
H  =  N \int dx^-dy^- \bigg[
 (g^2/4)  M_{--,ij}(x^-,y^-) |x^- - y^-|
M_{--,ij}(x^-,y^-) \\
   ~~~~~~~~~~~~~ -(g^2/4N) M_{--,ii}(x^-,x^-) |x^- - y^-|
M_{--,jj}(y^-,y^-) \\
   ~~~~~~~~~~~~~ -i (m^2/4)\;
{\rm sgn}(x^- - y^-) M_{--,ii}(x^-,y^-)\bigg]
\end{array}
\label{nine}
\end{equation}
The fermionic constraint (\ref{fivea}) reads in this language (after
we have put $A_-=0$):
\be
\begin{array}{l}
2i\del_{y^-} M_{++,ij} (x^-,y^-) + m M_{+-,ij} (x^-,y^-) = 0 \\
2i\del_{x^-} M_{+-,ij} (x^-,y^-) + m M_{--,ij} (x^-,y^-) = 0
\end{array}
\label{ninea}
\ee
Note that $N$ factors out of the hamiltonian and the subleading term
which arises because we are dealing with $SU(N)$
{\em rather than} $U(N)$ drops out in the large $N$ limit.

\vspace{3 ex}   %1 ex = height of the letter x

\noindent{\bf Fixing the Global Gauge Invariance}

\vspace{3 ex}

The choice of the gauge $A_- = 0$ still allows for global $SU(N)$
colour rotations, and in principle the wavefunction can carry a
representation of this global group or its subgroup depending on the
dynamical situation.  In 2-dimensional QCD if we restrict to finite energy
states then the linear Coulomb potential will ensure that coloured
asymptotic states do not appear.  We summarize this as a condition
that the wave function is invariant under global $SU(N)$ symmetry
\begin{equation}
\begin{array}{l}
[E^{ab}(x^-=+\infty) - E^{ab}(x^-=-\infty)] | \Psi \rangle \\
 = 2 \int dx^- \left( \psi^{\dagger b}_{i-}(x^-)
\psi^a_{i-} (x^-) - {1 \over
N} \delta^{ab} \psi^{\dagger c}_{i-} (x^-) \psi^c_{i-} (x^-) \right)
|\Psi\rangle = 0
\end{array}
\label{ten}
\end{equation}
The baryon number operator on the light-cone is defined by
\be
\begin{array}{l}
B_-^{(-)}=
{1 \over N} \sum^N_{a=1} \int dx^- \bar\psi^{+a}_{i\alpha} (x)
(\gamma^0\gamma^+)^{\alpha\beta}   \psi^a_{i\beta} (x)
\\
  ~~~~={1 \over N} \sum^N_{a=1} \int dx^- \bar\psi^{+a}_{i-} (x)
  \psi^a_{i-} (x)
\label{eleven}
\end{array}
\ee
The appearance of the $\gamma^0\gamma^+$ is due to the fact that
our baryon number is an integral on the light cone rather than
on a space-like surface. We use the superscript $(-)$ to
explicitly remind ourselves of this fact. The subscript says that
fermions of only the $-$ chirality appear in the integral.
One can show that this operator commutes with the hamiltonian
in \eq{nine}\ and so is conserved.

\vspace{3 ex}

\noindent{\bf Operator algebra of the string field}

\vspace{3 ex}

We can easily calculate the closed algebra satisfied by the string
field $M_{\alpha\beta,ij} (x,y)$ using the fermion anti-commutation
relation  \break $\{ \psi_-(x^-,x^+), \psi_-^\dagger(y^-, x^+) \} =
{1\over 2}\delta(x^- - y^-)$%
\footnote{The factor of ${1\over 2}$ is due to the factor of $2$
present in the kinetic term for $\psi_-$ in the action \eq{onea}.}.
Since in our case only the $M_{--}$ are dynamical we present
the algebra satisfied by them%
\footnote{Henceforth we shall drop the subscripts $_{--}$ from
$M$ since those will be the only ones we will be interested in (unless
otherwise stated).}:
\begin{equation}
\begin{array}{l}
\left[ M_{i_1j_1} (x^-_1,y^-_1;x^+), M_{i_2j_2}
(x^-_2,y^-_2;x^+)\right] \\
{}~~~~~~~~~~~~~~~~~=
{1 \over 2N}~\delta_{j_1i_2} \delta (y^-_1 -x^-_2)
M_{i_1j_2} (x^-_1,y^-_2;x^+) \\
{}~~~~~~~~~~~~~~~~~- {1 \over 2N}~\delta_{j_2i_1}
\delta (y^-_2 - x^-_1)~M_{i_2j_1} (x^-_2,y^-_1;x^+)
\end{array}
\label{twelve}
\end{equation}
We recognize (\ref{twelve}) as the infinite dim algebra
$\wf  \equiv
W_\infty \otimes U(n)$.  We recall that in the $c=1$
string field theory we had constructed the bilocal
operator $\Phi (x,y) =
\psi (x) \psi^+ (y)$ in terms of non-relativistic fermions which
satisfied the $W_\infty$ algebra \cite{COADJOINT}:
\begin{equation}
\left[ \Phi (x,y), \Phi (x', y')\right] = \delta (x' - y)~\Phi(x,y') -
\delta (y' - x) \Phi (y,x')
\end{equation}
There the analogue of the baryon number was the fermion number $N
=\int dx\; \psi^\dagger \psi.$  We mention here that $W_\infty \otimes
U(n)$ algebras have previously appeared in a different context in
\cite{BAKAS}.

\vspace{5 ex}

\noindent{\large\bf 2.
The co-adjoint orbit of $\wf = W_\infty \otimes U(n_f)$
and the classical action}

\vspace{5 ex}

We now show that the requirement of global colour invariance
(\ref{ten})
implies a quadratic constraint on $M_{ij} (x,y)$.
Using
fermion anti-commutation relation (at equal $x^+$)
we can prove the identity,
\be
\begin{array}{l}
\int dz^- M_{ik}(x^-,z^-)M^{kj}(z^-,y^-) \\
{}~~~= {1\over 2}M_{ij} (x^-,y^-) +
(1/N^2) \psi^a_{i-} (x^-) \psi^{\dagger
b}_{j-} (y^-) \int dz^- \sum_k
\psi^{a+}_{k-} (z^-) \psi^b_{k-} (z^-)
\end{array}
\end{equation}
Now using the requirement of global colour invariance (\ref{ten})
it is easy
to see that the following quadratic constraint is satisfied in the
physical space of states,
\be
M^2 = {1\over N^2} M \int dz^- \sum_k \psi^{\dagger c}_{k-}(z^-)
\psi^c_{k-}(z^-) + {1\over 2}M
\label{twelvea}
\ee
where we have used a compact matrix notation, $M_{ij}(x^-,y^-)$
being the $(ix^-, jy^-)$ element of a matrix $M$. This may
be rewritten as
\be
M^2 =  M({\over 2}1 + {1\over N} {\rm Tr}(1 -  M))
\ee
where the `Tr' refers to a combined flavour and `space' (in the sense
of $x^-$)  summation. The quantity ${\rm Tr}(1 -  M))$ is simply
the baryon number operator \eq{eleven}. Since it is conserved we may
simply replace it by its eigenvalue. We shall denote this by $B$,
that is
\be
{1 \over N} \sum^N_{a=1} \int dx \psi^{\dagger a}_{i-} (x) \psi^a_{i-}
(x) |\Psi\rangle
= {\rm Tr}( 1- M) |\Psi \rangle  =  B |\Psi\rangle
\label{fourteen}
\ee
The constraint now becomes
\begin{equation}
M^2 |\Psi\rangle = \left ({1\over 2}+ {B
\over N} \right) M |\Psi\rangle
\label{thirteen}
\end{equation}
We recall that the analogues of (\ref{thirteen})
and (\ref{fourteen}) have appeared previously
in the discussion of $c = 1$ field theory
\cite{COADJOINT}.  The classical analogues
of these constraints specify the co-adjoint orbit of $\wf = W_\infty
\otimes U(n)$, in the limit of large $N$.

Let us now construct the classical action in the large $N$ limit.  In
view of the constraints on $M$ the procedure is identical to that we
employed in $c=1$ field theory. The classical phase space is
constructed in terms of the expectation value $\langle M \rangle$ of
the operator $M$ in the coherent states of the algebra $\wf$.
Analogous to the operator constraint \eq{thirteena}\ on physical
states the classical phase space satisfies a constraint
\cite{PATHINTEGRAL}\ in terms of
the expectation value $\langle M \rangle$; it reads $\langle M
\rangle^2 = \langle M\rangle$.
Henceforth for simplicity of notation we will denote
$\langle M \rangle$ by $M$ itself.

The construction of the action now proceeds in exactly the same way as
was discussed for the $c = 1$ model \cite{COADJOINT,PATHINTEGRAL}.
Here we have the co-adjoint orbit of the group $\wf$ described by the
constraints
\be
\begin{array}{l}
M^2 = M \\
B = {\rm Tr}(1 - M)
\end{array}
\label{fifteen}
\ee
The path-integral can be constructed by regarding the four-fermi
interaction perturbatively. Like in \cite{PATHINTEGRAL}\ the
information about the {\em filling of the Fermi sea} is contained
in the choice of the particular co-adjoint orbit.
The action is given by
\begin{equation}
S = N\bigg[2i\int_\Sigma
dsdx^+\, {\rm Tr}(M \left[\del_+ M, \del_s M\right]) -
\int_{-\infty}^\infty dx^+  {\rm Tr}\big( {im^2\over 4} SM
+ {g^2\over 4}M \M\big)  \bigg]
\label{fifteena}
\end{equation}
where
\be
\M_{ij}(x^-,y^-) \equiv |x^- -y^-| M_{ij}(x^-,y^-),\; S(x^-,y^-)
= {\rm sgn}(x^- - y^-)
\label{fifteenb}
\ee
The region $\Sigma$ of
$(s,x^+) $ integration is  the lower half plane
$ x^+ \in (-\infty, +\infty), s \in (-\infty, 0] $ and we have
the boundary condition that $M(x^+,s=0) = M(x^+)$ and
$M(x^+, s) \to$ a constant ($x^+$-independent) matrix as $s\to
-\infty$.

The equation of motion can be derived by considering infinitesimal
motion on the co-adjoint orbit which
preserves the constraints (\ref{fifteen}).
Such a motion is given by $\delta M = i[\epsilon, M]$, which in
expanded form reads
\begin{eqnarray}
\delta M_{ij} (x^-,y^-) & = & i\int dx^-_1 \sum_{i_1}
\epsilon_{i,i_1} (x^-,x^-_1) M_{i_1,j} (x^-_1,y^-)
\nonumber  \\
& - & i\int dx^-_1 \sum_{i_1}
M_{i,i_1} (x^-,x^-_1) \epsilon_{i_1,j} (x^-_1,y^-)
\end{eqnarray}
By making such a variation in the action we can derive the following
equation of motion
\be
i\del_+ M  =  (im^2/8)[M,S]+ (g^2/4) \left[\M , M \right]  \\
\label{sixteen}
\ee

\vspace{5 ex}

\noindent{\large\bf
3. Classical Solution and Fluctuations: Representation of $W_{\infty+}
\otimes W_{\infty-}\otimes U(n_f)$}

\vspace{5 ex}

For simplicity we first consider the case
of one flavour. It is trivial to construct static ($x^+$-independent)
solutions which are translationally invariant, in other words solutions
of the form
\begin{equation}
\begin{array}{l}
M(x^-,y^-;x^+) = (1/2\pi) \int
dk_-  dk'_- M(k_-,k'_-;x^+) \exp(-ik_-x^- +  ik'_-y^-) \\
M(k_-, k'_-;x^+) =  f(k_-) \delta(k_- - k'_-)
\end{array}
\label{sixteena}
\end{equation}
The equation of motion (\ref{sixteen}) is automatically satisfied.
To satisfy the quadratic constraint (\ref{fifteen}) we need
\be
f(k_-)^2 =  f(k_-)
\label{seventeen}
\ee
The other (Baryon number)
constraint basically defines the filling of the Dirac sea.

\vspace{3 ex}

\noindent{\bf Vacuum Solution}

\vspace{3 ex}

Note that the solution
$ f(k_-) =  \theta(k_- - k_F) $
satisfies (\ref{seventeen}). Indeed this precisely corresponds to
the filled Dirac sea of free relativistic fermions in which the
Fermi level $k_F$ is determined by the Baryon number:
$ \int_{-\infty}^{k_F} dk_-  =  B$.  Clearly  $B$
is defined only through a regularization. Let us
choose a regularization such that the choice $k_F=0$ corresponds
to the regularized Baryon number = 0.

We therefore have a vacuum solution in the $B=0$ sector given by
\be
M_0(k_-,k'_-) = \theta (k_-) \delta(k_- - k'_-)
\label{eighteen}
\ee
We emphasize that though this solution corresponds to the filled
Dirac sea for a system of non-ineracting fermions this is indeed
a solution in our case of the fully interacting theory
(the interaction term in the equation of motion trivially
vanishes  for the condition (\ref{sixteena})).

\vspace{3 ex}

\noindent{\bf Fluctuations: Meson Spectrum as Representation of
$W_{\infty+}\otimes W_{\infty-}$}

\vspace{3 ex}

Let us parametrize  fluctuations around (\ref{eighteen}) as
\be
M = e^{iW/\sqrt{N}} M_0  e^{-iW/\sqrt{N}} =
M_0 + (i/\sqrt{N}) [ W, M_0] - (1/2N) [W,[W,M_0]] + o(N^{-3/2})
\label{nineteen}
\ee
where $e^{iW/\sqrt{N}}$ are arbitrary $\wf$-group elements. Note
that this parametrization automatically ensures that the
constraints are satisfied. We put in the factor of $\sqrt{N}$
explicitly to characterize fluctuations.

In the momentum-space notation (\ref{nineteen}) reads%
\footnote{In the following
we shall skip the subscript $_-$ from the momentum
labels for simplicity of typing.}
\be
\begin{array}{l}
M(k,k';x^+) = \theta(k) \delta(k-k') + (i/\sqrt{N})
(\theta(k') - \theta(k)) W(k,k';x^+) \\
 ~~~~~ - \int_{k''} (1/2N) (\theta(k) + \theta(k')
- 2 \theta(k'') ) W(k,k'';x^+) W(k'',k';x^+) + o(N^{-3/2})
\end{array}
\label{nineteena}
\ee
We see that upto linear order in \eq{nineteena}\ only the $W(k,k')$'s
with mixed signs of $k,k'$ appear. This leads to a
set of very interesting observations.

Let us introduce the following notation (the discussion in the rest of
the subsection is for a given $x^+$ which we do not write explicitly)
\be
\begin{array}{l}
W^{++}(k, k') =  W(k, k') \\
W^{+-}(k, k') =  W(k, - k') \\
W^{-+}(k, k') =  W(- k, k') \\
W^{--}(k, k') =  W(- k, - k') \\
\hbox{where}\;\; k, k' > 0
\end{array}
\ee
and similarly for the $M(k,k')$'s%
\footnote{The $\pm$ signs here denote the sign of momenta
and are not to be confused with chirality.}.

We note that the quantum algebra of the $M(k,k')$'s
\be
[M(k,k'), M(l,l')] = (1/2N)
\big( -\delta(k-l') M(l,k') + \delta(k'-l) M(k,l') \big)
\ee
leads to the following structure of commutation relation
between the $M^{\pm\pm}$'s
\be
[M^{++}(k,k'), M^{++}(l,l')] = (1/2N)
\big( -\delta(k-l') M^{++}(l,k') + \delta(k'-l) M^{++}(k,l') \big)
\label{comm1}
\ee
\be
[M^{--}(k,k'), M^{--}(l,l')] = (1/2N)
\big(- \delta(k-l') M^{--}(l,k') + \delta(k'-l) M^{--}(k,l') \big)
\label{comm1'}
\ee
\be
[M^{++}, M^{--}]=0
\label{comm1''}
\ee
\be
\begin{array}{l}
 ~[M^{++}(k,k'), M^{+-}(l,l')] = (1/2N)  \delta(k'-l) M^{+-}(k,l')  \\
 ~[M^{++} (k,k'), M^{-+} (l,l')] =- (1/2N)  \delta(k-l') M^{-+}(l,k')
\end{array}
\label{comm2}
\ee
\be
\begin{array}{l}
 ~[M^{--}(k,k'), M^{+-}(l,l')] = -(1/2N)  \delta(k-l') M^{+-}(l,k')  \\
 ~[M^{++}(k,k'), M^{-+}(l,l')] = (1/2N)  \delta(k'-l) M^{-+}(k,l')
\end{array}
\label{comm2'}
\ee
\be
\begin{array}{l}
 ~[M^{+-}(k,k'), M^{-+}(l,l')] = (1/2N)
\big( -\delta(k-l') M^{--}(l,k') +\delta(k'-l) M^{++}(k,l') \big) \\
  ~[M^{+-}, M^{+-}] = 0 = [M^{-+}, M^{-+}]
\end{array}
\label{comm3}
\ee
The commutation relations \eq{comm1},\eq{comm1'},\eq{comm1''}\ simply
state that there are two commuting sub-algebras of $\wf$ with
identical structure constants generated respectively by the $M^{++}$'s
and $M^{--}$'s.  Let's call these algebras $W_{\infty+}$ and
$W_{\infty-}$.  Let us imagine for the moment that in \eq{comm3}\
the generators $M^{++}$ and $M^{--}$ were replaced by their vacuum
expectation values. In that case the commutation relations
\eq{comm2},\eq{comm2'}, \eq{comm3}\ would simply mean that the
$M^{+-}$ and $M^{-+}$ are canonical conjugates which form
representation of the direct product of the `wedge' algebras
$W_{\infty+} \otimes W_{\infty+}$. We shall argue below that when
we use the expansion \eq{nineteena}\ in terms of the fluctuations
$W(k,k')$ in effect such a replacement of operators by their vacuum
expectation values can be done in the large $N$ limit.

Let us rewrite \eq{nineteena}\ in the $\pm$ notation. We get,
keeping upto the first non-trivial order in fluctuations
\be
\begin{array}{l}
M^{++}(k,k') =  \delta(k-k') - (1/N)\int_0^\infty  dk''
W^{+-}(k,k'')W^{-+}(k'',k')  + o(N^{-3/2})  \\
M^{+-}(k,k') = -(i/\sqrt{N})W^{+-}(k,k') + o(N^{-1}) \\
M^{-+}(k,k') = (i/\sqrt{N})W^{-+}(k,k') + o(N^{-1}) \\
M^{--}(k,k') = (1/N) \int_0^\infty dk'' W^{-+}(k,k'')W^{+-}(k'',k)
\end{array}
\label{fluct+-}
\ee

We find that the commutation relations for the $M$'s imply that
\be
\begin{array}{l}
  ~[W^{+-}(k,k'), W^{-+}(l,l')] = (1/2)\delta(k-l')\delta(k'-l) +
o(N^{-1/2})\\
  ~[W^{+-},W^{+-}]=[W^{-+},W^{-+}] = 0 + o(N^{-1/2})
\end{array}
\ee
and that
\be
\begin{array}{l}
 ~[{\mbar}^{++}(k,k'), W^{+-}(l,l')] = (1/2)\delta(k'-l) W^{+-}(k,l') +
o(N^{-1/2})\\
 ~[{\mbar}^{++}(k,k'), W^{-+}(l,l')] = - (1/2)\delta(k-l') W^{-+}(l,k')
+ o(N^{-1/2})
\end{array}
\ee
Here ${\mbar}^{++} \equiv N M^{++}$. Similar stataments are true for
$M^{--}$.
Note that it is the $\mbar$'s that in the large $N$-limit have
an $N$-independent structure constant.
Thus, in the limit $N=\infty$ we get a Heisenberg algebra
of the $W^{+-}, W^{-+}$ which forms a module of
$W_{\infty+}\otimes W_{\infty-}$.

In the following we shall analyze the equations of motion for the
fluctuations $W^{+-}, W^{-+}$  to identify an infinite tower of
mesons. Thus the Heisenberg algebra can essentially be identifed
with the algebra of canonical commutation relations of those fields.

\vspace{3 ex}

\noindent{\bf Spectrum}

\vspace{3 ex}

The action \eq{fifteena}\ based on $W^f_\infty$ coadjoint
orbit can be rewritten in terms of the fluctuations $W$ as%
\be
\begin{array}{l}
S = - 2\int dx^+ \int_0^\infty dk \int_0^\infty dk' \bigg[
W^{+-} (k',k;x^+)i\del_+ W^{-+}(k,k';x^+) \\
{}~- (m^2/ 4)(1/k + 1/k') W^{+-}(k,k';x^+) W^{-+} (k',k;x^+)
\\
 ~ + (g^2/8\pi) W^{+-} (k',k;x^+) \int_k^{-k'} (dp/p^2)
\big( W^{-+}(k-p,k'+p; x^+) - W^{-+} (k,k';x^+) \big) \\
 ~ + (g^2/8\pi) W^{-+} (k',k;x^+) \int_k^{-k'} (dp/p^2)
\big( W^{+-}(k'+p,k-p; x^+) - W^{+-} (k',k;x^+) \big)
\bigg] \\
 ~~~~~~~~~~~~~~~~~~~~~~~~~~~~+ 0(N^{-1/2})
\end{array}
\ee
This gives rise to the equation of motion
\be
\begin{array}{l}
i\del_+ W^{-+}(k, k'; x^+) =
 (m^2/ 4)(1/k + 1/k') W^{-+} (k,k';x^+)\\
 ~-(g^2/ 4\pi)\int_k^{-k'} dp
(1/p^2) \big[W^{-+}(k-p, k'+p;x^+) -
W^{-+}(k,k';x^+) \big] + o(N^{-1/2})
\end{array}
\label{twenty}
\ee
where $k$ and $k'$ are both $\ge 0$. Let us now define
\be
\begin{array}{l}
r_- =  k+ k' \\
x = k'/r_-
\end{array}
\label{twoone}
\ee
and also change the variable from $p$ to $y$ in the integral
on the r.h.s. of \eq{twenty}, where $y$ is defined by
\be
y = {p+k' \over r_-}
\ee
Clearly both $x$ and $y$ range over the interval $[0,1]$. Finally,
writing
\be
W^{-+}(k,k';x^+) = \int {dr^+ \over 2\pi} \phi(x;r_-,r_+) e^{ir_+x^+}
\ee
we get
\be
4r_- r_+ \phi(x) =  m^2( {1\over x} + {1\over 1-x} )
-{g^2\over \pi} \int_0^1 {dy \over (y - x)^2}
\big( \phi(y) - \phi(x))
\label{twotwo}
\ee
which is the same as the 'tHooft equation. This implies the
existence of an infinite tower of relativistic particle spectrum
\cite{THOOFT2}.

We therefore conclude that {\em in the large $N$ limit the infinite
tower of mesons in two-dimensional QCD ({\em
the higher string modes}) form a representation of the
algebra $W_{\infty+} \otimes W_{\infty-} $.}

We remark that although the above algebra is represented in the space
of fluctuations it does not commute with the hamiltonian.  One
interesting consequence of this, as seen by considering the action of
the $\mbar^{++}$ or $\mbar^{--}$'s on the $\phi(x; r_-,r_+)$'s, is
that {\em different mass levels transform into one another under the
action of this algebra}. This is indeed as one would expect in a
string field theory.

\vspace{3 ex}

\noindent{\bf Generalization to many flavours}

\vspace{3 ex}

The above discussion has a simple generalization to $n_f > 1$. The
vacuum solution is
\be
M_{0,ij}(k,k') = \delta_{ij}\theta(k) \delta(k-k')
\label{manyflavour}
\ee
The `wedge' subalgebra's are generated by $M^{++}_{ij}(k,k')$ and
$M^{--}_{ij}(k,k')$.  The `dynamical variables' are  $W^{+-}_{ij}(k,k')$
and $W^{-+}_{ij}(k,k')$. There is a 'tHooft equation for each
hermitian generator of the $U(n_f)$ group. Thus we have a tower of
meson corresponding to each generator. In other words, the
small fluctuations around the vacuum \eq{manyflavour}\ {\em
form a repressentation of $W_{\infty+} \otimes W_{\infty-} \otimes
U(n_f)$.}

\vspace{5 ex}

\noindent{\large\bf 4.  Concluding Remarks}

\vspace{5 ex}

We make here a few observations about solitonic configurations in the
theory. In the large $N$-limit the string field $M_{ij}(x^-,y^-)$
becomes classical which essentially means that the quartic fermion
interaction in \eq{six}\ can be considered equivalent to a quadratic
potential term:
\be
\begin{array}{l}
H  =  (g^2/4) \int dx^-dy^- \bigg( \psi^a_{i-} (x^-)
\psi^{a+}_{j-} (y^-) V_{ij}(x^-,y^-) \\
{}~~~~ - {1 \over N}~\psi^a_{i-}(x^-)
\psi^{+a}_{i-}(x^-) V_{jj}(y^-,y^-) \bigg)\\
 ~~~~V_{ij}(x^-,y^-) \equiv \M^c_{ij}(x^-,y^-)=
|x^- - y^-| M^c_{ij}(x^-,y^-)
\end{array}
\label{Hartree}
\end{equation}
Here the superscript $^c$ on $M$ denotes `classical' solution.  In the
previous section we have discussed in detail a particular classical
solution in the translationally invariant sector. It is clear that
every classical solution of the equation of motion \eq{nineteen}\ for
$M$ would correspond to some `Hartree-Fock' potential $V$. The
fermions moving in these potentials are like quasi-particles of Landau
theory of fermi liquids. We can try to construct solutions of the
many-body problem where we place $N$ quarks in a single localized
quasi-particle state above a filled sea of quasi-particle levels.
Such classical solutions would correspond to baryons. The
quantitative details of this picture will shortly appear elsewhere
\cite{BARYON}. The
point that we want to make here is that we can again
go through arguments similar to those used in the previous
section to construct a representation of $W_\infty^f$ around
each of these classical solutions $M^c_{ij}(x^-,y^-)$. Thus
we find that {\em solitons in `string theories' are characterized
by infinite-dimensional groups like $W_\infty^f$ which are
generated by non-local loop operators and are therefore capable
of describing an infinite-parameter space of deformation of
these solitons.}

We end with a few remarks comparing QCD$_2$ strings with continuum
strings.  As we remarked in the introduction, there are many
differences {\sl a priori.} In the matrix model formulation of
continuum strings, the double scaling limit ensures that a smooth
continuum limit of the triangulated surfaces can be taken. No such
facility exists in the QCD$_2$ case in any obvious sense, the problem
being compounded by the fact that in two dimensions
the triple-gluon vertex, which is
crucial in forming a surface, vanishes in an axial gauge. One possible
reconciliation could be that in two dimensions the metric
fluctuations on the world sheet for an open string theory essentially
boil down to fluctuations of the length of the string. Presumably the
meson spectrum reflects just this vibrating degree of freedom
\cite{BARS}.

\vspace{3 ex}

\noindent{\bf Acknowledgements:}  We would like to thank
R. Shankar for participation in the early stages of this work.

%\vspace{5 ex}
\newpage

\begin{center}
{\large\bf  Appendix}
\end{center}

\vspace{5 ex}

\noindent We summarize our notation for the light-cone
variables:
\be
\begin{array}{l}
{\rm coordinates}:  x^\pm \equiv x^0 \pm x^1   \\
{\rm metric}: ds^2 = (dx^0)^2 - (dx^1)^2 = dx^+dx^-; \;
g_{+-}= g_{-+}={1\over 2}  \\
{\rm gauge\, fields}:  A_\pm = (\del x^0/ \del x^\pm) A_0
+ (\del x^1/ \del x^\pm) A_1 = {1\over 2} (A_0 \pm A_1) \\
{\rm momenta:}   k_\pm = {1\over 2} (k_0 \pm k_1) \\
\hbox{mass-shell:}  4 k_+k_- = m^2 \\
\hbox{gamma-matrices:} \gamma^\pm = \gamma^0 \pm \gamma^1
\end{array}
\label{three}
\ee
We use the following explicit representation for gamma-matrices:
\be
\gamma^0=\sigma^1, \; \gamma^1= -i \sigma^2, \; \gamma^5\equiv
\gamma^0\gamma^1 = \sigma^3
\ee
\be
\gamma^+ = \left( \begin{array}{cc}
		   0 & 0 \\
		   2 & 0  \end{array} \right) ; \;
\gamma^- = \left( \begin{array}{cc}
		   0 & 2 \\
		   0 & 0  \end{array} \right)
\ee
We also use the following notation for
the Dirac fermion $\psi$:
\be
\psi = \left( \begin{array}{c}
		\psi_-  \\
		\psi_+  \end{array}   \right)
\label{sixa}
\ee

\end{document}